\documentclass[aps,prb,amsfonts,amsmath,amssymb,twocolumn]{revtex4}
\usepackage{bm}
\usepackage{graphicx}
\usepackage{amsmath}
\usepackage{amstext}
\usepackage{amssymb}
\usepackage{amsfonts}
\usepackage{amsbsy}
\usepackage{verbatim}
\usepackage{wasysym}

\def\vr{\bar{r}}
\def\vR{\bar{R}}
\def\bx{\boldsymbol{x}}
\def\by{\boldsymbol{y}}
\def\ba{\boldsymbol{a}}

\def\br{\boldsymbol{r}}

\def\bk{\boldsymbol{k}}

\def\bq{\boldsymbol{q}}
\def\bQ{\boldsymbol{Q}}

\def\bN{\boldsymbol{N}}

\def\bz{\boldsymbol{z}}

\def\rSU{{\rm SU}}

\def\bJ{\boldsymbol{J}}

\def\balpha{\boldsymbol{\alpha}}
\def\bsigma{\boldsymbol{\sigma}}

\def\rSU2{{\rm SU}(2)}

\def\bI{\boldsymbol{I}}
\def\bmu{\boldsymbol{\mu}}
\def\bnu{\boldsymbol{\nu}}
\def\uonexy{{\rm U}(1)_{XY}}

\begin{document}

\title{Non-abelian descendant of abelian duality in a two-dimensional frustrated quantum magnet}
\author{Michael Hermele}
\affiliation{Department of Physics, University of Colorado, Boulder, Colorado 80309, USA}

\date{\today}
\begin{abstract}
Several recent works on quantum criticality beyond the Landau-Ginzburg-Wilson paradigm  have led to a number of field theories, potentially important for certain two-dimensional magnetic insulating systems, where criticality is not very well understood.  This situation highlights the need for non-perturbative information about criticality in two spatial dimensions (three space-time dimensions), which is a longstanding challenge.  As a step toward addressing these issues, we present evidence that the O(4) vector model is dual to a theory of Dirac fermions coupled to both ${\rm SU}(2)$ and ${\rm U}(1)$ gauge fields.  Both field theories arise as low-energy, long-wavelength descriptions of a frustrated XY model on the triangular lattice.  Abelian boson-vortex duality of the lattice model, together with the emergence of larger non-abelian symmetry at low energies, leads to this rare example of duality in two spatial dimensions involving non-abelian global symmetry and fermions, but without supersymmetry.  The duality can also be viewed as a bosonization of the Dirac fermion gauge theory.
\end{abstract}
\maketitle

\section{Introduction}
\label{sec:intro}

Recent theoretical progress on quantum criticality in two spatial dimensions ($d = 2$) has elucidated a number of critical phenomena beyond the Landau-Ginzburg-Wilson (LGW) paradigm, particularly in the context of $d=2$ magnetic insulators.\cite{rantner01, rantner02, senthil04a, senthil04b}  In the LGW paradigm, one identifies one or more order parameters and uses them  to construct an effective field theory based on symmetry alone, and studies criticality in this theory.  The field theories that arise outside the LGW framework make up a much broader class than those arising within it, often involve fermions and/or gauge fields, and in general are less understood than those theories arising within the LGW framework.

Most analytical understanding of quantum criticality in $d=2$ is based on  perturbative calculations in large-$n$ and $d = 3 - \epsilon$ expansions, and it would be extremely useful to obtain non-perturbative information that reduces our reliance on these formal limits.  This is even true  within the LGW framework, although, for example, the ${\rm O}(n)$ model is relatively well understood after extensive numerical and analytical studies.\cite{kleinertbook}  The field theories that arise beyond the LGW paradigm have been much less studied, and non-perturbative information is more urgently needed.  Toward this end, in this paper I argue that the O(4) model (with some anisotropy terms) is identical at low energies to an apparently quite different theory involving fermions and gauge fields, which is closely related to field theories describing one of the main classes of non-LGW criticality, namely critical spin liquids.  This relation is a rare example of duality in $D = d+1 = 3$ space-time dimensions involving non-abelian global symmetries, non-abelian gauge fields, and fermions, without supersymmetry.

The proposed relation, which is the main result of this paper, is that the ${\rm O}(4)$ vector model is dual to a theory of $N_f = 2$ four-component Dirac fermions coupled to both ${\rm SU}(2)$ and ${\rm U}(1)$ gauge fields.  We dub the latter theory QCED3, as it is in a sense a combination of $D=3$ quantum electrodynamics (QED3) and quantum chromodynamics (QCD3) with $N_c = 2$ colors.  To obtain this relation, we begin with an XY antiferromagnet on the triangular lattice.  This model can be treated directly by a Landau theory approach, and this leads to the O(4) field theory, with various anisotropy terms breaking the O(4) symmetry down to that of the original model.  Alternatively, using abelian boson-vortex duality,\cite{peskin78, dasgupta81} the XY model is mapped onto a dual lattice model, from which we obtain QCED3 as a low-energy effective description.  The non-abelian duality between the O(4) model and QCED3 thus emerges at low energy as a descendant of abelian boson-vortex duality on the lattice.

This result builds in a crucial way on work of Alicea \emph{et. al.}, where it was proposed, beginning with the same frustrated XY model, that the ${\rm O}(4)$ vector model is dual to $N_f = 2$ two-component Dirac fermions coupled to a ${\rm U}(1)$ Maxwell gauge field \emph{and} a ${\rm U}(1)$ Chern-Simons gauge field.\cite{alicea05a}  Later, this result was obtained from a different point of view by Senthil and Fisher, who also argued that the ${\rm O}(4)$ model with a topological term ($\theta$-term at $\theta = \pi$) is dual to the same ${\rm U}(1)$ gauge theory but with no Chern-Simons term.\cite{senthil06}  In these examples, either the original or dual theory involves a topological term, which prevents the development of a controlled large-$n$ or $d = 3 - \epsilon$ expansion.  Because such expansions are possible for the dual partners without topological terms, these results are very useful for understanding their partners that do have topological terms.  However, because only one dual partner can be directly analyzed, such dualities do not lead to much additional understanding of the critical behavior itself.  The result presented here differs in the crucial respect that \emph{both} the original and dual theories lack topological terms, and thus both admit controlled expansions, potentially allowing greater insight into the critical properties.

In general, two field theories are dual when: (1) the two theories have identical low-energy physics, and (2) the variables in the two theories are \emph{mutually nonlocal}.  That is, the fields representing one theory cannot be written as a local function of the fields in the other, and vice versa.  The latter condition excludes trivial, local changes of variables.  The classic example of such a duality for $D= 3$ is between the XY model and the abelian Higgs model.\cite{peskin78,dasgupta81}  In general, it is known how to construct explicit duality relations for models with abelian global symmetries and abelian gauge fields -- without fermions -- in arbitrary space-time dimension.\cite{savit80}  However, as in the field theories of interest here, many of the most interesting cases involve non-abelian global symmetries, non-abelian gauge fields and/or fermions, where no general techniques exist to construct useful duality relations.  The primary exceptions are in $D = 2$, where bosonization\cite{gogolin-book} can be thought of as a duality, and in a number of supersymmetric field theories in higher dimensions.\cite{seiberg94a, seiberg94b, intriligator96, kapustin99}  The results of this paper are a step toward understanding duality in a less restrictive context.

It is important at this stage to make a distinction between duality of field theories and duality of critical fixed points.  By duality of field theories, we shall mean that as the parameters of the two theories are varied, they have the same phases and critical points, and all the same low-energy, long-wavelength behavior.  Duality of fixed points, on the other hand, is a stronger statement, and is most easily defined by discussing its meaning in the context of the O(4)-QCED3 duality.  The O(4) model has a Wilson-Fisher fixed point which can be accessed in a controlled fashion by a large-$n$ expansion, where one considers the ${\rm O}(n)$ model.  Also, QCED3 has a critical fixed point that can be accessed in the large-$N_f$ expansion; we assume that this critical point survives down to the value $N_f = 2$ of interest here.  By duality of fixed points, we would mean in this case that the O(4) Wilson-Fisher fixed point is identical to the QCED3 fixed point.  However, it must be emphasized that this statement need not hold for the two \emph{field theories} to be dual.  It could be the case that the O(4) and QCED3 fixed points are distinct, and, in each of the two field theories, both fixed points are present. 
The O(4)-QCED3 duality as proposed is to be understood primarily as a duality between field theories.  Whether it also leads to a duality between the O(4) and QCED3 fixed points is a more challenging question, and the available evidence  does not allow us to reach a firm conclusion.  However, it is plausible that the two fixed points are identical.

We now briefly outline the remainder of the paper.  In Sec.~\ref{sec:prelims} we introduce the frustrated triangular lattice XY model, and review the treatment of Ref.~\onlinecite{alicea05a} up to the point where it deviates from our approach.  In particular, we discuss a direct Landau theory approach, which leads to the O(4) field theory (Sec.~\ref{sec:directxy}), and the dual lattice model resulting from the boson-vortex duality transformation (Sec.~\ref{sec:dualmodel}).  Next, we describe our route from the dual model to a continuum field theory (Sec.~\ref{sec:vortexfrac}), and proceed to give a more precise statement of the proposed duality  (Sec.~\ref{sec:statement}).  In the following two sections, we present the principal evidence for the duality:  In Sec.~\ref{sec:evidence1} we set up a dictionary between operators in the O(4) and QCED3 field theories, and in Sec.~\ref{sec:evidence2} we demonstrate that the three stable phases of the original XY model can be realized in QCED3.  We conclude in Sec.~\ref{sec:discussion} with a discussion of open issues and possible directions for future work.  Three appendices contain various technical details.

\section{Preliminaries and Precursors}
\label{sec:prelims}

\subsection{Direct analysis of XY model}
\label{sec:directxy}

The initial steps needed to obtain the ${\rm O}(4)$-QCED3 duality have already been carried out by Alicea \emph{et. al.},\cite{alicea05a} so we review their results here and in Sec.~\ref{sec:dualmodel}, while introducing notation that will be important later on.  At a certain point, which we identify in Sec.~\ref{sec:dualmodel}, our treatment deviates from theirs.  As we shall very briefly outline, following their route leads instead to the duality between the ${\rm O}(4)$ model and Dirac fermions coupled to both Maxwell and Chern-Simons ${\rm U}(1)$ gauge fields.  Our route to the ${\rm O}(4)$-QCED3 duality is described beginning in Sec.~\ref{sec:vortexfrac}.

\begin{figure}
\includegraphics[width=3in]{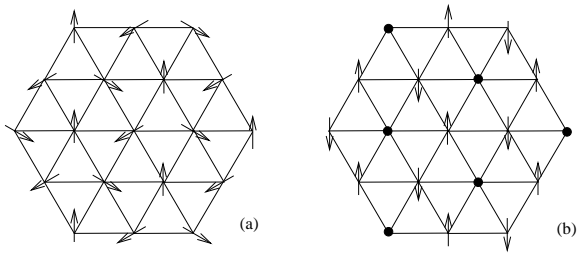}
\caption{Depiction of the coplanar $120^\circ$ ordered state (a), and one of the two collinear states (b).  The filled circles represent sites with zero ordered moment.  The other collinear state is depicted in Ref.~\onlinecite{alicea05a}.}
\label{fig:orders}
\end{figure}

The starting point, here and in Ref.~\onlinecite{alicea05a}, is an XY antiferromagnet on the triangular lattice, whose Hamiltonian is
\begin{equation}
\label{eqn:xymodel}
{\cal H} = U \sum_{\br} n_{\br}^2
+ J \sum_{\langle \br \br' \rangle} \cos (\phi_{\br} - \phi_{\br'} ) \text{.}
\end{equation}
Here, $J > 0$ and the second sum is over nearest-neighbor bonds.  On every site of the triangular lattice $\br$ there is a ${\rm U}(1)$ quantum rotor, with angular position $e^{i \phi_{\br}}$ and integer-valued angular momentum $n_{\br}$, which satisfy the commutation relation $[n, e^{i \phi}] = e^{i \phi}$.    This can be thought of as an effective theory for a $S = 1$ XY antiferromagnet.  Alternatively, by the usual quantum-classical correspondence, its partition function describes a classical XY antiferromagnet of stacked triangular lattices, which has been extensively studied\cite{kawamura98, pelissetto02, delamotte04, calabrese04} (see also Ref.~\onlinecite{alicea05a} for a more detailed overview of the literature).
In this context, it is known that the model can realize three phases (provided next-neighbor ferromagnetic exchange is included), which are: (1) paramagnet (2) coplanar 120$^\circ$ magnetic order (Fig.~\ref{fig:orders}) and (3) two different collinear magnetic orders (Fig.~\ref{fig:orders} and Ref.~\onlinecite{alicea05a}).  The paramagnet, coplanar state, and one of the two collinear states meet at a multicritical point (upon tuning both $J$ and the further-neighbor exchange), which is the Wilson-Fisher fixed point of the ${\rm O}(4)$ vector model.  Either collinear state can arise adjacent to the multicritical point, depending on the sign of a dangerously irrelevant 6th-order term in the Landau theory (discussed below), and the distinction between the two collinear states will not play an important role in the present discussion.

The symmetries of Eq.~(\ref{eqn:xymodel}) will play a crucial role in our analysis, so we enumerate them here.  The triangular lattice space group is generated by a translation by one lattice constant in the $x$-direction ($T_x$), a counterclockwise rotation by $\pi/3$ about a lattice site ($R_{\pi/3}$), and a reflection ${\cal R}_y : (x,y) \to (x, -y)$.  The operators $n_{\br}$ and $\phi_{\br}$ transform as scalars under the space group.  There is also a charge-conjugation (or spin flip) symmetry
\begin{eqnarray}
{\cal C} &:& n_{\br} \to - n_{\br} \\
{\cal C} &:& \phi_{\br} \to - \phi_{\br} \text{.}
\end{eqnarray}
And the antiunitary time-reversal operation
\begin{eqnarray}
{\cal T} &:& n_{\br} \to -n_{\br} \\
{\cal T} &:& \phi_{\br} \to \phi_{\br} + \pi \text{.}
\end{eqnarray}  
Finally there is the ${\rm U}(1)$ phase rotation, which sends $\phi_{\br} \to \phi_{\br} + \alpha$.  To distinguish it from other symmetry groups that will arise later on, we shall refer to this symmetry as ${\rm U}(1)_{XY}$.

To expose the ${\rm O}(4)$ structure, one goes to a path integral description of the partition function in terms of $\phi_{\br}(\tau)$, where $\tau$ is imaginary time, and focuses on those configurations of the phase field with lowest action.  These are given by $e^{i \phi_{\br}} \sim e^{i \bQ \cdot \br} z_1(\br, \tau) + e^{- i \bQ \cdot \br} z_2(\br, \tau)$, where $\bQ = (4\pi/3)\bx$ (the lattice constant is set to unity), and $z_{1,2}$ are slowly varying functions of $\br$ and $\tau$.  The vectors $\pm \bQ$ lie at the corners of the hexagonal Brillouin zone and are the ordering wavevectors of the $120^\circ$ state.  The action on $z_{1,2}$ of the microscopic symmetries is enumerated in Appendix~\ref{app:symms}.

Constrained by the microscopic symmetries, one then writes a continuum effective Lagrange density
\begin{equation}
\label{eqn:o4-lagrangian}
{\cal L} = | \partial_{\mu} Z |^2 + r Z^\dagger Z + u (Z^\dagger Z)^2 + v_4 |z_1|^2 |z_2|^2 \text{,}
\end{equation}
where terms of order $|z|^6$ and higher have been discarded, and $Z^T = (z_1 \, z_2)$.  For $v_4 = 0$, the model's continuous symmetry is ${\rm SO}(4)$, which is broken down to ${\rm U}(1) \times {\rm U}(1)$ for $v_4 \neq 0$.  The 6th order terms break the continuous symmetry down to ${\rm U}(1)_{XY}$.  The $v_4$ and $r$ terms are relevant perturbations to the ${\rm O}(4)$ multicritical point, while the allowed 6th order (and higher) terms are irrelevant.  At the mean-field level, the coplanar state arises when $r < 0$ and $v_4 > 0$, and has $\langle z_1 \rangle \neq 0$ and $\langle z_2 \rangle = 0$ (or vice-versa).  The collinear state obtains for $v_4 < 0$ and has $| \langle z_1 \rangle| = |\langle z_2 \rangle| \neq 0$.  Depending on the relative phase of $z_1$ and $z_2$ two different collinear ordering patterns are possible; this phase is determined by the sign of a dangerously irrelevant 6th order term.

Some of the properties of the ${\rm O}(4)$ critical point are known from numerical simulations and high-order perturbative calculations; see Ref.~\onlinecite{kleinertbook} for an extensive discussion (and tabulation) of critical properties of ${\rm O}(n)$ models.  Approximately, the critical exponent $\nu \approx 0.75$, and $\eta \approx 0.027$.  These exponents can be translated into scaling dimensions of fields by $\operatorname{dim} Z = (1 + \eta)/2 \approx 0.51$ and $\operatorname{dim} Z^\dagger Z = 3 - 1/\nu \approx 1.67$.

It will be convenient to use the fact that ${\rm SO}(4) \simeq [ {\rm SU}(2) \times {\rm SU}(2)]/ Z_2$.  We thus introduce the matrix
\begin{equation}
{\cal Z} = \left( \begin{array}{cc}
z_1 & -z^*_2 \\
z_2 & z^*_1
\end{array} \right) \text{.}
\end{equation}
A general ${\rm SO}(4)$ rotation is realized by ${\cal Z} \to U_L {\cal Z} U_R$, where $U_{L,R} \in {\rm SU}(2)$.  We shall thus refer to left and right ${\rm SU}(2)$ rotations, denoted by ${\rm SU}(2)_{L,R}$, with conserved currents $\bJ^{L,R}_{\mu}$.  In terms of the fields,
\begin{eqnarray}
\bJ^L_{\mu} &=& \frac{1}{2} \operatorname{tr} \Big[ {\cal Z}^\dagger \bsigma (\partial_{\mu} {\cal Z}) \Big] \\
\bJ^R_{\mu} &=& \frac{1}{2} \operatorname{tr} \Big[ \bsigma {\cal Z}^\dagger  (\partial_{\mu} {\cal Z}) \Big]  \text{,}
\end{eqnarray}
where $\bsigma = (\sigma^1, \sigma^2, \sigma^3)$ is a vector of the $2\times 2$ Pauli matrices.  The ${\rm U}(1)_{XY}$ symmetry is a subgroup of ${\rm SU}(2)_R$ and is generated by $2 (\bJ_0^R)^z$, while $\bJ^L_0$ generates rotations of the two slowly varying fields $z_1$ and $z_2$ into one another.

One of the major pieces of evidence for the ${\rm O}(4)$-QCED3 duality will be a dictionary identifying operators in the two field theories.  To that end, we now enumerate some important operators in the ${\rm O}(4)$ model.  Operators can be labeled by $(\ell_L , \ell_R)$, the total angular momentum quantum numbers of ${\rm SU}(2)_L \times {\rm SU}(2)_R$.  It will sometimes be useful to also specify $(m_L, m_R)$, the projection quantum numbers of the angular momenta along the $z$-axis.  Note that $2 m_R$ is the ${\rm U}(1)_{XY}$ charge. We have already discussed the boson field $Z$ itself, which transforms as $(1/2, 1/2)$, and the currents $\bJ^L_{\mu}$ and $\bJ^R_{\mu}$, which transform as $(1,0)$ and $(0,1)$, respectively, and are both vectors under Lorentz rotations (\emph{i.e.} space-time rotations).  We shall also consider the following 9 Lorentz scalar bilinears in $Z$:
\begin{eqnarray}
\bN &=& Z^\dagger \bsigma Z \\
\bI &=& Z^T (i \sigma^2) \bsigma Z \\
\bI^* &=& (\bI)^* \text{.}
\end{eqnarray}
Together these make up a $\ell_L = \ell_R = 1$ multiplet.  We shall also consider the operators ${\cal O}_r = Z^\dagger Z$ and ${\cal O}_v = |z_1|^2 |z_2|^2 - (1/6) (Z^\dagger Z)^2$.  ${\cal O}_r$ has $\ell_L = \ell_R = 0$, and ${\cal O}_v$ is a member of a $\ell_L = \ell_R = 2$ multiplet with $m_L = m_R = 0$.  These play an important role as they are the terms in the Lagrangian that must be tuned to reach the ${\rm O}(4)$ critical point.  Next, defining the ${\rm O}(4)$ vector $\phi_i$ ($i = 1,\dots,4$) by $z_1 = \phi_1 + i \phi_2$ and $z_2 = \phi_3 + i \phi_4$, we define the operator ${\mathfrak C} = i \epsilon_{i j k l} \epsilon_{\mu \nu \lambda} \phi_i \partial_{\mu} \phi_j \partial_{\nu} \phi_k \partial_{\lambda} \phi_l$, which is odd under ${\cal T}$ and ${\cal R}_y$.  In a nonlinear sigma model version of the O(4) field theory, where the constraint $Z^\dagger Z = 1$ is imposed, if ${\mathfrak C}$ is integrated over space-time and added to the action it becomes the topological $\theta$-term.\cite{abanov00}  While ${\mathfrak C}$ does not have any special topological significance as a local operator, we shall refer to it as the topological density.

\subsection{Dual model}
\label{sec:dualmodel}
\begin{figure}
\includegraphics[width=3in]{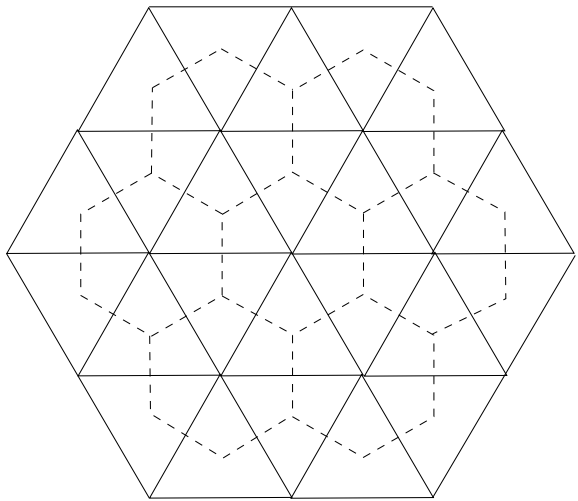}
\caption{Relation between the dual honeycomb lattice (dashed lines) and the triangular lattice of the original XY model (solid lines).  Honeycomb lattice sites correspond to triangular lattice plaquettes.}
\label{fig:duallattice}
\end{figure}

In Ref.~\onlinecite{alicea05a}, boson-vortex duality was applied to the Hamiltonian Eq.~(\ref{eqn:xymodel}).  This is a straightforward procedure for such a lattice XY model; here we review some crucial aspects, and more details are found in Ref.~\onlinecite{alicea05a}.  The degrees of freedom of the dual model are vortices, which are bosons residing on the sites of the dual honeycomb lattice (Fig.~\ref{fig:duallattice}), and the vector potential and electric field of a non-compact ${\rm U}(1)$ gauge field, which reside on the nearest-neighbor bonds of the honeycomb lattice.  Crucially, the vortices are at half-filling (an average of one-half vortex per honeycomb lattice site).  This is a direct consequence of the frustration in the original XY model:  the XY exchange term of Eq.~(\ref{eqn:xymodel}) can be rewritten as $- J \sum_{\langle \br \br' \rangle} \cos(\phi_{\br} - \phi_{\br'} + {\cal A}_{\br \br'} )$, where the flux associated with the non-fluctuating vector potential ${\cal A}_{\br \br'}$ is half a flux quantum, or $\pi$, for each triangular plaquette.  This background flux has the consequence of forcing in half a vortex per site (on average) in the dual theory.

For technical convenience, we shall take the vortices to experience a hardcore repulsion, and only allow zero or one vortex on each honeycomb site.  Such a modification of microscopic parameters will not affect the universal low-energy properties of phases and critical points in which we shall be interested.

The dual Hamiltonian is
\begin{eqnarray}
\label{eqn:dual-hamiltonian}
{\cal H}_{{\rm dual}}
 &=& U' \sum_{\hexagon} \big[ (\nabla \times a)_{\hexagon} \big]^2 + J' \sum_{\langle \vr \vr' \rangle} e^2_{\vr \vr'} \nonumber \\
&-& t_v \sum_{\langle \vr \vr' \rangle} \big[ e^{i a_{\vr \vr'}} v^+_{\vr} v^-_{\vr'} + \text{H.c.} \big] \text{.}
\end{eqnarray}
Here, $\bar{r}$ labels the sites of the dual honeycomb lattice.  $e_{\bar{r} \bar{r}'}$ and $a_{\bar{r} \bar{r}'}$ are, respectively, the electric field and vector potential residing on the honeycomb link joining $\bar{r}$ and $\bar{r}'$.  On the same link these satisfy the canonical commutation relation $[e, a] = i$.  The first term is a sum over all  hexagonal plaquettes, and $(\nabla \times a)_{\hexagon}$ is the lattice curl, which is the discrete line integral of $a_{\bar{r} \bar{r}'}$ taken around the perimeter of a given hexagonal plaquette.  The latter two terms are sums over nearest-neighbor honeycomb links.  Because the vortices are hardcore bosons, their creation, destruction and number operators are represented using the $S = 1/2$ spin operators $(v^x_{\bar{r}}, v^y_{\bar{r}}, v^z_{\bar{r}})$, satisfying the usual commutation relation $[v^i_{{\bar r}}, v^j_{\bar{r}'} ] = i \delta_{\bar{r} \bar{r}'} \epsilon^{i j k} v^k_{\bar{r}}$. A vortex at site $\bar{r}$ is created by $v_{\bar{r}}^+ = v^x_{\bar{r}} + i v^y_{\bar{r}}$ and destroyed by $v^-_{\bar{r}} = (v^+_{\bar{r}})^\dagger$.  The vortex number operator is $N_v(\bar{r}) = v^z_{\bar{r}} + 1/2$.  The Hamiltonian is supplemented with the Gauss' law constraint
\begin{equation}
(\operatorname{div} e)_{\bar{r}} = v^z_{\bar{r}} \text{,}
\end{equation}
where $(\operatorname{div} e)_{\bar{r}} = \sum_{\bar{r}' \sim \bar{r}} e_{\bar{r} \bar{r}'}$ is the lattice divergence of $e_{\bar{r} \bar{r}'}$.
The parameters of the dual Hamiltonian are $U' \simeq U$, $J' \simeq J$, and the vortex hopping $t_v$.  It should be emphasized that Eq.~(\ref{eqn:dual-hamiltonian}) is not an exact rewriting of the original XY model, and instead should be thought of as a low-energy effective theory.  Conversely, the XY model of Eq.~(\ref{eqn:xymodel}) can \emph{also} be thought of as a low-energy effective theory for the dual model Eq.~(\ref{eqn:dual-hamiltonian}).

It is important to spell out the connection between the ${\rm U}(1)$ gauge field and the original XY model degrees of freedom.  The magnetic flux represents the $\uonexy$ charge density,
\begin{equation}
\label{eqn:flux-charge}
\frac{1}{2\pi} (\nabla \times a)_{\hexagon} \sim n_{\br},
\end{equation}
where $\hexagon$ is the honeycomb hexagon surrounding the triangular lattice site $\br$.  Also, the electric field is related to the $\uonexy$ current by
\begin{equation}
\sin (2 \pi e_{\bar{r} \bar{r}'} ) \sim \sin (\phi_{\br} - \phi_{\br'}) \text{,}
\end{equation}
where $(\br, \br')$ is the unique triangular lattice bond crossing the honeycomb link $(\bar{r}, \bar{r}')$.  In the dual representation, $\uonexy$ charge conservation is represented as the conservation of magnetic flux.  Moreover, insertion of $\uonexy$ charge (as by acting with $e^{i \phi_{\br}}$) is represented by the insertion of a quantized $2\pi$ magnetic flux.  Such flux insertions are space-time magnetic monopole events, and will play a crucial role in our analysis.

Under the space group symmetry, $e_{\bar{r} \bar{r}'}$ and $a_{\bar{r} \bar{r}'}$ transform as pseudovectors, and $v^z_{\bar{r}}$ as a pseudoscalar.  $v^{\pm}_{\bar{r}}$ transforms as a scalar under $T_x$ and $R_{\pi/3}$, and ${\cal R}_y : v^{\pm}_{\bar{r}} \to v^{\mp}_{\bar{r}'}$, where $\bar{r}'$ is the image of $\bar{r}$ under the reflection.  Under charge-conjugation both $e$ and $a$ are odd, while
\begin{eqnarray}
{\cal C}: v^z_{\bar{r}} &\to& - v^z_{\bar{r}} \\
{\cal C}: v^{\pm}_{\bar{r}} &\to& v^{\mp}_{\bar{r}} \text{.}
\end{eqnarray}
Finally, under time-reversal $a$ is odd while $e$ is even, and
\begin{eqnarray}
{\cal T}: v^z_{\bar{r}} &\to&  v^z_{\bar{r}} \\
{\cal T}: v^{\pm}_{\bar{r}} &\to& v^{\pm}_{\bar{r}} \text{.}
\end{eqnarray}

The challenge at this stage is to use the dual Hamiltonian to construct a continuum low-energy effective theory, which would then provide a dual description of the ${\rm O}(4)$ critical point.  A natural approach would be to begin with Eq.~(\ref{eqn:dual-hamiltonian}) and construct a functional integral in terms of the phase field of the vortices $v^+ \sim e^{i \theta}$.  However, as the vortices are at half-filling, the resulting functional integral is plagued with Berry phase terms, and does not directly lead to a useful continuum limit.  Instead, it is necessary to choose some other set of variables that is amenable to a continuum description.  

Alicea \emph{et. al.} chose to represent the vortices in terms of fermions using statistical transmutation, with the price of introducing a Chern-Simons gauge field $\alpha_{\bar{r} \bar{r}'}$.\cite{alicea05a}  In a flux-smearing mean-field treatment, the vortices are fermions at half-filling on the honeycomb lattice (with zero background magnetic flux), and thus have a massless Dirac dispersion.    To include fluctuations about the mean-field state, one re-couples the Dirac fermions to both the Chern-Simons gauge field and the Maxwell gauge field $a$ that arose in the boson-vortex duality.  The resulting Euclidean Lagrangian is
\begin{eqnarray}
{\cal L} &=& \bar{\Psi} \gamma_{\mu} (\partial_{\mu} + i a_{\mu} + i \alpha_{\mu} ) \Psi +
\frac{1}{2 e^2} \sum_{\mu} (\epsilon_{\mu \nu \lambda} \partial_{\nu} a_{\lambda} )^2 \nonumber \\
&+& \frac{i}{4\pi} \alpha_{\mu} \epsilon_{\mu \nu \lambda} \partial_{\nu} \alpha_{\lambda} + \cdots \text{.}
\end{eqnarray}
The first term is the kinetic energy of the fermions and contains the coupling to the two gauge fields.  The second term dictates that the $a_{\mu}$ gauge field obeys Maxwell dynamics, and the last term is the Chern-Simons term for $\alpha_{\mu}$.  The coefficient of the Chern-Simons term is precisely that needed to attach $2\pi$ flux and transmute fermions into bosons (and vice versa).  The ellipsis represents other (important) perturbations consistent with the underlying microscopic symmetries.  

While this field theory is an intriguing dual representation of the ${\rm O}(4)$ model, the presence of the Chern-Simons term seriously hinders any direct analysis of it.  It was conjectured in Ref.~\onlinecite{alicea05a} that the Chern-Simons term can simply be dropped without affecting the critical properties, but the arguments in favor of this conjecture are not conclusive, and are questionable given the later results of Ref.~\onlinecite{senthil06}.  Motivated in part by a desire to avoid these issues, here we pursue a different approach that also leads to a fermionic gauge theory representation of the ${\rm O}(4)$ model, but without any topological terms.

\section{Duality between ${\rm O}(4)$ model and QCED3.}

\subsection{Vortex Fractionalization Route to Dual Effective Theory}
\label{sec:vortexfrac}

The present approach also begins with the dual Hamiltonian Eq.~(\ref{eqn:dual-hamiltonian}).  The challenge is to construct a continuum effective theory that deals with the half-filling of the vortices, but also avoids the difficulties associated with statistical transmutation and the Chern-Simons term.  To do this, we look to the theory of spin liquids in $S = 1/2$ Heisenberg models, which are after all also models of half-filled hardcore bosons.  One route to describe spin liquids, in particular critical spin liquids that lack a spin gap, is to formally represent hardcore bosons (or spins) as bilinears of fermionic slave particles.\cite{wen02}  Therefore it is reasonable to hope that a similar approach can describe criticality -- in particular, the ${\rm O}(4)$ critical point -- in the present model, and we shall argue that this is indeed the case. \footnote{It is natural to ask whether it is useful to instead split the vortices into \emph{bosonic} slave particles.  A cursory investigation suggests that this route does not lead to an interesting dual description of the ${\rm O}(4)$ critical point.}

We represent the vortices using the fermionic operators $f_{\bar{r} \alpha}$, where $\alpha = 1,2$:
\begin{eqnarray}
v^+_{\bar{r}} &=& f^\dagger_{\bar{r}1} f^{\vphantom\dagger}_{\bar{r} 2} \\
v^z_{\bar{r}} &=& \frac{1}{2} ( f^\dagger_{\bar{r}1} f^{\vphantom\dagger}_{\bar{r}1} - f^\dagger_{\bar{r}2} f^{\vphantom\dagger}_{\bar{r}2} ) \text{.}
\end{eqnarray}
With the local constraint $f^\dagger_{\bar{r} \alpha} f^{\vphantom\dagger}_{\bar{r} \alpha}  =1$ this change of variables provides an exact rewriting of the model.  In these variables there is a local ${\rm SU}(2)$ gauge redundancy,\cite{affleck88a, dagotto88} which can be exposed by defining
\begin{equation}
\psi_{\bar{r}} = \left(
\begin{array}{c}
f^\dagger_{\bar{r} 1} \\
\epsilon_{\bar{r}} f_{\bar{r} 2} 
\end{array} \right) \text{,}
\end{equation}
where $\epsilon_{\bar{r}} = 1$ ($-1$) for $\bar{r}$ in the A (B) sublattice.  The fermions satisfy the local constraint equations $\psi^\dagger_{\bar{r}} \mu^i \psi_{\bar{r}} = 0$, where $\mu^i$ ($i = 1,2,3$) are the $2 \times 2$ Pauli matrices.  This is simply the condition that the ${\rm SU}(2)$ gauge charge is zero at every lattice site.  The vortex operators $v^{\pm}_{\bar{r}}$ and $v^z_{\bar{r}}$ can be written in manifestly ${\rm SU}(2)$ gauge-invariant forms, and so all physical operators are gauge invariant.

To proceed, we pass to a functional integral representation, where the action is $S = S_{{\rm U}(1)} + S_f$, where
\begin{eqnarray}
\label{eqn:u1-micropart}
S_{{\rm U}(1)} &=& \frac{1}{4 J'} \int d\tau \sum_{\langle \vr \vr' \rangle} \big[ \partial_{\tau} a_{\vr \vr'} 
- ( \Delta a_0)_{\vr \vr'} \big]^2 \nonumber \\
&+& U' \int d\tau \sum_{\hexagon} (\nabla \times a)^2_{\hexagon} \text{.}
\end{eqnarray}
Here, we have introduced the notation $(\Delta f)_{\vr \vr'} = f(\vr') - f(\vr)$ for a lattice derivative.
The fermionic part of the action is 
\begin{eqnarray}
S_f &=& \int d\tau \sum_{\vr} \psi^\dagger_{\vr} \Big[ \partial_{\tau} + \frac{i}{2} a_0(\vr,\tau) 
+ \frac{i \alpha_0^i(\vr, \tau) \mu^i}{2} \Big] \psi^{\vphantom\dagger}_{\vr} \nonumber \\
&-& \frac{i}{2} \int d\tau \sum_{\vr} a_0(\vr, \tau) \nonumber \\
&+& t_v \int d\tau \sum_{\langle \vr \vr' \rangle} \Big[ e^{i a_{\vr \vr'}} v^{+}_{\vr} v^{-}_{\vr'} + \text{H.c.} \Big] \text{.}
\label{eqn:f-micropart}
\end{eqnarray}
Note that we have made the gauge transformation $a_{\bar{r} \bar{r}'} \to a_{\bar{r} \bar{r}'} + \pi$ to change the sign of the last term.  We have also introduced the Lagrange-multiplier field $\alpha^i_0$ ($i = 1,2,3$), which enforces the ${\rm SU}(2)$ gauge constraint and can be thought of as the time-component of the ${\rm SU}(2)$ gauge field.

The last term of Eq.~(\ref{eqn:f-micropart}) is quartic in the fermion operators, and to arrive at a candidate low-energy effective theory it can be decoupled using a Hubbard-Stratonovich field residing on the bonds of the lattice.  One searches for mean-field saddle points of this field, and each such saddle point (upon including often important fluctuations) leads to a low-energy effective theory.  Many distinct effective theories can be generated in this way,\cite{wen02} and one of the challenges of this approach is to decide which theory (if any) accurately captures the physics of the model at hand.  In the present case we are guided by the requirement that the effective theory should be able to reproduce the phases and critical points that are known to be present from the analysis of the original XY model.

Rather than carry out the above mean-field procedure explicitly, we shall simply guess the form of the low-energy theory.  To do this we write down an effective lattice gauge theory that reduces to the above model in a particular limit.  This is equivalent to choosing  a particular mean-field saddle point and then including the fluctuations about it.  The effective lattice theory is obtained by replacing $S_f$ with
\begin{eqnarray}
S'_f &=& \int d\tau \sum_{\vr} \psi^\dagger_{\vr} \Big[ \partial_{\tau} + \frac{i}{2} a_0(\vr,\tau) 
+ \frac{i \alpha_0^i(\vr, \tau) \mu^i}{2} \Big] \psi^{\vphantom\dagger}_{\vr} \nonumber \\
&-& \frac{i}{2} \int d\tau \sum_{\vr} a_0(\vr, \tau) \nonumber \\
&+& t \int d\tau \sum_{\langle \vr \vr' \rangle} \Big[ e^{- i a_{\vr \vr'} / 2} \psi^\dagger_{\vr} U_{\vr \vr'} \psi^{\vphantom\dagger}_{\vr'} + \text{H.c.} \Big] \text{.}
\label{eqn:f-effective}
\end{eqnarray}
Here $U_{\vr \vr'}$ is the spatial part of the ${\rm SU}(2)$ gauge field.  It should be noted that, because the gauge field $a_{\vr \vr'}$ is noncompact, it is perfectly legitimate to have the object $e^{i a_{\vr \vr'}/2}$ appearing in the action.
We also include a Maxwell action $S_g$ for the ${\rm SU}(2)$ gauge field, with overall strength proportional to the coupling constant $1/g^2$.  When $g \to \infty$ and $t$ is small, $U_{\vr \vr'}$ can be integrated out perturbatively in $t$.  At leading order ($t^2$), one recovers the original dual Hamiltonian given by the action $S = S_{{\rm U}(1)} + S_f$, with an additional nearest-neighbor repulsive interaction between vortices.  This interaction is not important for our purposes, as it does not change the symmetry of the model and is not expected to affect its universality class.

On the other hand, if we take $g$ small and $U'$ large, then the fluctuations of both gauge fields are suppressed, and in an appropriate gauge $U_{\vr \vr'} \approx 1$ and $\alpha^i_0 \approx a_{\vr \vr'} \approx a_0 \approx 0$.  In this mean-field limit the fermions are described by the Hamiltonian
\begin{equation}
{\cal H}_{{\rm MFT}} = t \sum_{\langle \vr \vr' \rangle} \big[ \psi^\dagger_{\vr} \psi_{\vr'} + \text{H.c.} \big] \text{,}
\end{equation}
which is simply nearest-neighbor hopping of half-filled fermions on the honeycomb lattice.  The corresponding low-energy theory is given by focusing on the excitations near the Dirac nodes and reintroducing the coupling to the gauge fields.  The resulting Lagrangian density is
\begin{eqnarray}
{\cal L}_{{\rm QCED3}} &=& \bar{\Phi} \Big[ - i \gamma_{\mu} \big( \partial_{\mu} + \frac{i a_{\mu}}{2} + \frac{i \alpha^i_{\mu} \mu^i}{2} \big) \Big] \Phi \\
&+& \frac{1}{2 e^2} \sum_{\mu} (\epsilon_{\mu \nu \lambda} \partial_{\nu} a_{\lambda})^2
+ \frac{1}{2 g^2} f^i_{\mu \nu} f^i_{\mu \nu} \nonumber \text{.}
\end{eqnarray}

Here we have introduced the continuum Dirac field $\Phi$, which is related to $\psi_{\bar{r}}$ as discussed in Appendix~\ref{app:fermion-contlimit}.  $\Phi$ is an eight-component object;   these eight components arise from the two-component nature of $\psi_{\bar{r}}$,  the two bands needed to represent each Dirac node, and the two-component flavor index corresponding to the two distinct nodes in the Brillouin zone.  It is useful to define three different sets of Pauli matrices acting in this 8-component space; each set corresponds to its own type of ${\rm SU}(2)$ rotations.  The $\mu^i$ Pauli matrices act in the ${\rm SU}(2)$ gauge space and generate gauge transformations.  The $\tau^i$ Pauli matrices act in the band index, or Lorentz,  space, and generate Lorentz transformations.  Finally, the $\nu^i$ Pauli matrices act in the ${\rm SU}(2)$ flavor space and generate flavor rotations.  $\Phi$ resides in the tensor product of these three ${\rm SU}(2)$ spaces, and products of different types of Pauli matrices (which commute) should be understood as matrix tensor products.  The action of the various Pauli matrices on $\Phi$ is given explicitly in Appendix~\ref{app:fermion-contlimit}.  It is convenient to think of $\Phi$ as composed of $N_f = 2$ flavors of four-component Dirac fermions, where each flavor transforms as a doublet under ${\rm SU}(2)$ gauge rotation and Lorentz transformations.  Both flavors carry the same ${\rm U}(1)$ charge of $1/2$ under the dual gauge field $a_{\mu}$.

We have also introduced the field strength of the ${\rm SU}(2)$ gauge field,
\begin{equation}
f^i_{\mu \nu} = \partial_{\mu} \alpha^i_{\nu} -  \partial_{\nu} \alpha^i_{\mu} + \epsilon_{i j k} \alpha^j_{\mu} \alpha^k_{\nu} \text{,}
\end{equation}
and the matrices $\gamma_{\mu}$ are defined in terms of the $\tau^i$ Pauli matrices as
\begin{equation}
\gamma_{\mu} = (\tau^3, \tau^2, -\tau^1) \text{.}
\end{equation}
Finally, we have defined
\begin{equation}
\bar{\Phi} = i \Phi^\dagger \tau^3 \text{.}
\end{equation}

A number of theories similar to ${\cal L}_{{\rm QCED3}}$, also involving Dirac fermions coupled to gauge fields, are interesting in the context of non-LGW criticality.  This interest stems from the fact that the microscopic symmetries can (in some cases) be enough to forbid the addition of any relevant perturbations to the action, and the fixed point thus describes a stable critical phase.\cite{rantner01, rantner02, hermele04}  Such critical phases have been discussed in a variety of physical settings.\cite{affleck88, marston89, franz01,vafek02, alicea05b, alicea06,ryu07, ran07,kaul08, cenkexu08a}

The global symmetries of ${\cal L}_{{\rm QCED3}}$ play a crucial role in our discussion. Aside from $D =3$ Poincar\'{e} invariance, the continuous global symmetry is ${\rm SU}(2) \times {\rm U}(1)$.  The global ${\rm SU}(2)$ consists of rotations between the two flavors of Dirac fermions generated by the $\nu^i$ Pauli matrices.  We thus dub it ${\rm SU}(2)_F$, and it has the conserved current
\begin{equation}
\bJ^F_{\mu} = \bar{\Phi} \gamma_{\mu} \bnu \Phi \text{.}
\end{equation}
The global ${\rm U}(1)$ is simply the $\uonexy$ symmetry.  Its realization in ${\cal L}_{{\rm QCED3}}$ follows directly from the boson-vortex duality transformation, and in particular the identification of $a_{\mu}$ magnetic flux and $\uonexy$ charge density [Eq.~(\ref{eqn:flux-charge})].  The associated conserved current is thus
\begin{equation}
j^G_{\mu} = \frac{1}{2\pi} \epsilon_{\mu \nu \lambda} \partial_{\nu} a_{\lambda} \text{,}
\end{equation}
the flux of the $a_{\mu}$ gauge field.

The field theory ${\cal L}_{{\rm QCED3}}$, like other $D = 3$ theories of massless Dirac fermions coupled to gauge fields, is solvable in the large-$N_f$ limit.\cite{appelquist86, rantner01, rantner02, hermele04, hermele05, ran06}  When $N_f \to \infty$ the fluctuations of the gauge fields are suppressed, and for most purposes the physics is identical to that of non-interacting fermions.  Expanding about this limit, correlation functions can be calculated order-by-order in $1/N_f$, and it is found that various operators acquire anomalous dimensions.  The principal result, then, is that the large-$N_f$ expansion describes an interacting critical fixed point,\cite{rantner01, rantner02} which we assume survives down to the case of interest, $N_f = 2$.   Below, we shall refer to the $N_f = 2$ incarnation of the large-$N_f$ fixed point as the QCED3 fixed point, which should be distinguished, of course, from the QCED3 field theory.

\subsection{Statement of the proposed duality}
\label{sec:statement}

A brief statement of the proposed duality is that the O(4) model and QCED3 possess identical low-energy physics.  Below, we shall elaborate on the meaning of this statement in order to give a more precise statement of the duality.  We then outline the approach underlying the evidence for the duality, which is described in the following sections.

It is useful to remark that both the O(4) model and QCED3, provided arbitrary perturbations consistent with the underlying microscopic symmetries are added to each, are expected to be valid low-energy effective descriptions of the original model.  This is expected simply because both were derived from the same microscopic starting point.  This means, in particular, that any phase or critical point of either the direct or dual low-energy theory is a phase or critical point that presumably exists somewhere in the parameter space of the original XY model.  It is not immediately obvious, however, that the O(4) model and QCED3 describe phases and critical points in the same part of parameter space.  This is the crucial fact that needs to be established for the duality to hold.

With this remark in mind, the proposed duality can be precisely stated as follows:  Beginning with ${\cal L}_{{\rm QCED3}}$, by adding operators consistent with the underlying microscopic symmetries one can tune the theory to an O(4) critical point identical to the Wilson-Fisher fixed point of the O(4) model.  We shall call the resulting fine-tuned Lagrangian ${\cal L}^c_{{\rm QCED3}}$.  Moreover, the effective Lagrangian
\begin{equation}
{\cal L}_{{\rm eff}} = {\cal L}^c_{{\rm QCED3}} + r \tilde{{\cal O}}_r  + v \tilde{{\cal O}}_v  \text{,}
\end{equation}
where $\tilde{{\cal O}}_r$ and $\tilde{{\cal O}}_v$ are QCED3 operators  identified below in Sec.~\ref{sec:evidence1}, is identical at low energy to the O(4) model Lagrangian of Eq.~(\ref{eqn:o4-lagrangian}).  These statements essentially amount to asserting that, indeed, the O(4) model and QCED3 describe the same region of the original model's parameter space.

Using the terminology introduced in Sec.~\ref{sec:intro}, this is a duality between field theories.  It is natural to ask if  there is \emph{also} a duality between the O(4) Wilson-Fisher fixed point and the QCED3 critical point.  That is, are the O(4) and QCED3 fixed points identical or distinct?  
 We do not have a definitive answer to this question, but we will see it is plausible that the two fixed points are dual.  This is discussed further in Sec.~\ref{sec:discussion}. 
Most of the evidence presented in Secs.~\ref{sec:evidence1} and~\ref{sec:evidence2} pertains to the duality of field theories.  However, we do present some information on scaling dimensions from the large-$N_f$ expansion for the QCED3 fixed point.  These results are not relevant for establishing the duality between field theories, but they do provide some information about a potential duality of fixed points.

In Sec.~\ref{sec:evidence1} we set up a dictionary of operators between QCED3 and the O(4) model.  In order to do this, it will be helpful to exploit the large-$N_f$ understanding of the QCED3 fixed point; this provides a controlled understanding of the field content of QCED3, which is our primary concern in establishing the duality between field theories.  Next, in Sec.~\ref{sec:evidence2}, we describe how to access the stable phases of the original XY model in the QCED3 description.

\section{Evidence for the duality: operator dictionary}
\label{sec:evidence1}

To construct a dictionary between QCED3 and O(4) model operators, we begin by identifying the continuous global symmetries of the two field theories.  We identify the ${\rm SU}(2)_F$ flavor symmetry of QCED3 with ${\rm SU}(2)_L$ of the O(4) model, which means that the currents $\bJ^F_{\mu}$ and $\bJ^L_{\mu}$ should also be identified.  
Next, we identify $j^G_{\mu}$ with $2 (\bJ^R_{\mu})^z$, as these are the realizations of the $\uonexy$ conserved current in QCED3 and the O(4) model, respectively.  

As QCED3 does not have manifest ${\rm SU}(2) \times {\rm SU}(2)$ symmetry, it is less clear how to identify the remaining components of $\bJ^R_{\mu}$.  An important part of the proposed duality  is that, at the ${\rm O}(4)$ critical point, the ${\rm SU}(2)_L \times {\rm U}(1)_{XY}$ global symmetry of QCED3 is enlarged to ${\rm SU}(2)_L \times {\rm SU}(2)_R$.  So, in particular, $\uonexy$ is enlarged to ${\rm SU}(2)_R$, which happens quite explicitly in the O(4) model.  Moreover, it should not be surprising that it is possible to enlarge ${\rm SU}(2) \times {\rm U}(1)$ to ${\rm SU}(2) \times {\rm SU}(2)$ by a suitable tuning of parameters, since both groups at least have the same Cartan subalgebra.  We shall identify candidate QCED3 partners of the remaining components of $\bJ^R_{\mu}$ below.

To continue setting up our dictionary, it will be convenient to break QCED3 operators into two classes.  The first class consists of all operators carrying zero $\uonexy$ charge, or, equivalently, zero magnetic flux of the dual gauge field $a_{\mu}$.  We refer to operators in the first class as non-monopole operators.  The second class contains all monopole operators, which do carry an $a_{\mu}$ flux.  The flux is quantized in multiples of $2\pi$, simply because the $\uonexy$ charge is quantized in the original XY model.  Non-monopole operators can be easily represented in terms of the fermion and gauge fields.  On the other hand, as is typical for topological disorder operators, monopole operators are more difficult to represent in terms of the fields of the theory.  However, they can be constructed using the state-operator correspondence of conformal field theory,\cite{borokhov02} and we shall take advantage of this approach here.

To be identified, two operators certainly must transform identically under Lorentz transformations, ${\rm SU}(2)_L$ (or, equivalently, ${\rm SU}(2)_F$), and $\uonexy$.  For non-monopole operators, in each case we have also verified, using the results enumerated in Appendix~\ref{app:symms}, that each pair of identified operators transforms identically under all the microscopic symmetries.  For monopole operators, on the other hand, it is only known how to \emph{partially} determine the action of microscopic symmetries, up to a few unknown parameters.\cite{alicea05a, alicea08, hermele08}  Here we shall do this, following the approach of Ref.~\onlinecite{hermele08}.  In each case the transformations of the corresponding O(4) model operator can be determined completely using the results of Appendix~\ref{app:symms}, and are consistent with the partial results for the QCED3 counterpart. 

We now proceed to identify some of the important O(4) model operators with QCED3 counterparts.

{\bf O(4) bilinear $\bN = Z^\dagger \bsigma Z$.}  This is perhaps the simplest operator to identify with a QCED3 partner.  We identify $\bN$ with
\begin{equation}
\tilde{\bN} = -i \bar{\Phi} \bnu \Phi \text{.}
\end{equation}
Both $\bN$ and $\tilde{\bN}$ transform as a vector under ${\rm SU}(2)_L$, a scalar under Lorentz rotation, and carry zero $\uonexy$ charge. 

In the limit $N_f \to \infty$, the scaling dimension $\Delta_{\tilde{\bN}}$ of $\tilde{\bN}$ approaches 2.  An inspection of the diagrams involved in the $1/N_f$ correction shows that the contributions from the ${\rm U}(1)$ and ${\rm SU}(2)$ gauge fields\cite{rantner02, ran06} simply add together, and
\begin{equation}
\Delta_{\tilde{\bN}} = 2 - \frac{128}{3 \pi^2 N_f} + {\cal O}(1/N_f^2) \text{.}
\end{equation}
While this result is not to be trusted quantitatively, the qualitative trend that $\Delta_{\tilde{\bN}} < 2$ is believed to be reliable.\cite{rantner02, hermele-erratum07}  This should be compared to the fact that, in the large-$n$ expansion of the ${\rm O}(n)$ model, $\Delta_{\bN} = 1 + {\cal O}(1/n)$.  It is reasonable that these scaling dimensions become the same for the case of interest ($n = 4$ and $N_f = 2$).

{\bf O(4) field $Z$.}  Since $Z$ carries $\uonexy$ charge of unity, it is a monopole operator carrying $2\pi$ flux.  To construct $2\pi$-monopoles in QCED3, we need to make use of the state-operator correspondence following Ref.~\onlinecite{borokhov02}.  Accessible treatments of the state-operator correspondence can be found in Refs.~\onlinecite{polchinskibook} and~\onlinecite{metlitski08}.    Briefly, for a $D=3$ Lorentz and scale invariant theory, such as the QCED3 fixed point, the state-operator correspondence states that local operators of the field theory are in one-to-one correspondence with quantum states of the same theory quantized on the two-dimensional unit sphere.  Scaling operators (\emph{i.e.} eigenoperators of a renormalization group transformation) are mapped to eigenstates of the Hamiltonian on the 2-sphere, and the scaling dimension of the operator is equal to the energy of the corresponding state.  

Here we are primarily concerned with using the state-operator correspondence to \emph{construct} monopole operators, by constructing the corresponding eigenstates of the Hamiltonian on the 2-sphere.  This can be done for QCED3 in the large-$N_f$ limit, where the fluctuations of both gauge fields are suppressed and the $N_f \to \infty$ Hamiltonian is simply that of $N_f$ flavors of non-interacting 4-component fermions on the unit sphere.  As $N_f$ is reduced from infinity down to the case of interest $N_f = 2$, gauge fluctuations will modify the states and change their energies (and hence their scaling dimensions).  
However, even for $N_f = 2$, the states constructed with gauge fluctuations suppressed still correspond to local operators (although no longer to scaling operators), and we still use them to understand the field content of $N_f = 2$ monopole operators.  

We shall see below that it will be useful to think in terms of $N_{f 2} = 2 N_f$ two-component fermions, each of which is a Lorentz doublet.  A monopole operator with flux $2\pi q$ corresponds to a state of these fermions with a background $a_{\mu}$ flux of $2 \pi q$ on the sphere.  Since the fermions carry $a_{\mu}$ charge of $1/2$, they feel only a flux of $\pi q$.  

Let us now consider a $2\pi$-flux monopole operator, where the fermions feel a flux of $\pi$ from $a_{\mu}$; this violates the Dirac quantization condition and is thus apparently problematic.  To illustrate the problem, recall that the
monopole's gauge field can be represented using a Dirac string carrying the $2\pi$-flux away from the center of the monopole in an infinitesimally thin solenoid.  However, since the fermions feel only a $\pi$-flux from the solenoid, it is a physical object, and the resulting object is thus not a local operator.  One might try to eliminate the Dirac string using the mathematical technology of sections and the monopole harmonics of Wu and Yang,\cite{wu76} but Dirac's quantization condition still enters there as a requirement that the transition function be single valued, and the problem is not avoided.

However, the Dirac quantization condition \emph{can} be repaired in the present case if flux in the ${\rm SU}(2)$ gauge field is also present, and so this must happen in order to get a local operator with unit $\uonexy$ charge.  In particular, we can consider putting a $2\pi$-flux monopole in $\alpha^3_{\mu}$ -- half of the 2-component fermions feel this as $\pi$ flux, and the other half feel it as $-\pi$ flux.  (Other strengths of monopoles can also be considered, but these either lead to states that will have higher energy -- and thus higher scaling dimension -- or continue to violate the Dirac quantization condition.)  Therefore, combined with the overall $\pi$-flux coming from $a_{\mu}$, there will be $N_f$ 2-component fermions moving in a background $2\pi$ flux, and $N_f$ fermions moving in zero flux.  An issue that immediately arises is that the resulting state is not invariant even under global ${\rm SU}(2)$ gauge transformations.  We return to this below, after discussing the structure of the state in the fixed gauge where there is a monopole in $\alpha^3_{\mu}$ and $\alpha^1_{\mu} = \alpha^2_{\mu} = 0$.  It should be noted that this configuration of the ${\rm SU}(2)$ gauge field is not itself topologically stable, but is induced by the topologically stable monopole in the ${\rm U}(1)$ gauge field.

To understand the structure of the $2\pi$-monopole states and thus the corresponding operators, we shall need the spectrum of a single 2-component Dirac fermion on the 2-sphere, moving in a uniform background flux $2\pi f$.  The energy levels for $|f| > 1$ are\cite{borokhov02}
\begin{equation}
E_p = \pm  \sqrt{p^2 + p |f|} \text{,}
\end{equation}
where $p$ is a nonnegative integer.  For $f = 0$ the form of the spectrum is the same, but $p$ is restricted to be positive and thus there are no $E_p = 0$ states. These levels are $(2 j_p + 1)$-fold degenerate, where $j_p = (1/2)(|f|  -1) + p$, and transform in the $(2 j_p + 1)$-dimensional representation of the ${\rm SU}(2)$ Lorentz group (\emph{i.e.} rotation symmetry of the 2-sphere).

\begin{figure}
\includegraphics[width=3in]{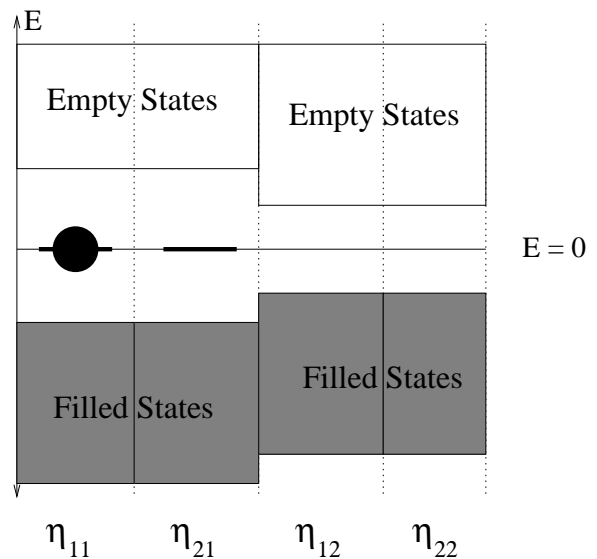}
\caption{Illustration of the $2\pi$-monopole state whose corresponding operator is the QCED3 partner of   the ${\rm O}(4)$ field $Z$.  The spectrum of each 2-component fermion $\eta_{a \alpha}$ is shown.  The fermions $\eta_{1 1}$ and $\eta_{21}$ both feel a net $2\pi$ flux through the sphere.  The energy level spectrum for each of these fermions is symmetric about zero energy, where for each fermion there is a single zero energy state.  The $\eta_{12}$ and $\eta_{22}$ fermions feel no flux, and also have a spectrum symmetric about zero energy, but with no zero energy state.  All negative energy levels are filled, and one of the two zero mode states is filled to obtain a state with zero total charge under both the $a_{\mu}$ and $\alpha^3_{\mu}$ gauge fields.}
\label{fig:zeromode}
\end{figure}

We now specialize to $N_f = 2$, the case of interest.  We represent $\Phi$ in terms of $N_{f 2} = 4$ two-component fermions $\eta_{a \alpha}$, where $a, \alpha = 1,2$ and $\Phi^T = (\eta^T_{1 1}, \eta^T_{1 2}, \eta^T_{2 1}, \eta^T_{2 2} )$.  The $a$ index transforms as a doublet of ${\rm SU}(2)_L$, and the $\alpha$ index as a doublet of ${\rm SU}(2)$ gauge rotations.  This means that the $\alpha = 1$ ($\alpha = 2$) fermions carry $\alpha^3_{\mu}$ charge of $1/2$ ($-1/2$).  Then $\eta_{1 1}$ and $\eta_{2 1}$ feel flux $f = 1$, while $\eta_{1 2}$ and $\eta_{2 2}$ feel $f = 0$.  The spectrum is as illustrated in Fig.~\ref{fig:zeromode}.

 In order for the state to correspond to a gauge-invariant operator, it must carry zero $a_{\mu}$ charge.  Moreover, we shall see below that it must also carry zero $\alpha^3_{\mu}$ charge.  Now, both $a_{\mu}$ and $\alpha^3_{\mu}$, and hence the corresponding charges, change sign under the charge conjugation symmetry ${\cal C}$, which acts as a particle-hole transformation on the $\eta$-fermions.  Crucially, ${\cal C}$ acts trivially on the $\alpha$ index of $\eta_{a \alpha}$ (see Appendix~\ref{app:symms}), so it does not exchange fermions feeling flux with those feeling no flux.  Acting with ${\cal C}$ on the state illustrated in Fig.~\ref{fig:zeromode} results in another state of the same schematic form -- the only change is in the occupation of the zero mode states -- which must therefore have the same charges as the original state.  The only consistent possibility is that both $a_{\mu}$ and $\alpha^3_{\mu}$ charges are zero, as needed.  Other states with the correct charges, and higher energy, can be obtained from this state by moving fermions between levels.
  
Now we return to the question of gauge invariance of the $2\pi$-flux monopole state.  To correspond to a gauge-invariant local operator, the state should be invariant under general ${\rm U}(1)$ and ${\rm SU}(2)$ gauge transformations.  We can construct a gauge-invariant state starting with the gauge-fixed state discussed above, denoted by $| \psi_0 \rangle$.  (Which of the two zero modes is filled is not important for this discussion, so we just focus on a single state.)  The gauge-invariant state is constructed by integrating over all possible gauge transformations:
\begin{equation}
\label{eqn:ginvt-monopole}
|\psi\rangle = \int [d G_1] \int [d G_2] G_1 G_2 | \psi_0 \rangle \text{,}
\end{equation}
where $G_1$ and $G_2$ are unitary operators implementing ${\rm U}(1)$ and ${\rm SU}(2)$ gauge transformations, respectively, and the integrals are taken over all such unitary transformations.  This state $| \psi \rangle$ is clearly gauge invariant, but we need to check that it does not vanish.  This is done in Appendix~\ref{app:nonzero}, where it is shown that zero $a_{\mu}$ and $\alpha^3_{\mu}$ charge is a necessary and sufficient condition for $| \psi \rangle$ not to vanish.

We are now in a position to discuss the quantum numbers of the $2\pi$-flux monopole states depicted in Fig.~\ref{fig:zeromode}, and thus those of the corresponding operator.  The negative energy single-particle levels, when fully occupied, are invariant under both Lorentz rotation and ${\rm SU}(2)_L$ flavor.  However, the fermions feeling $f=1$ have $2$ zero-energy Lorentz-singlet levels, which are filled with a single fermion.  This implies that the overall state transforms as a Lorentz singlet, but as a doublet under ${\rm SU}(2)_L$.  We thus denote the corresponding monopole operator as the two-component object $\tilde{Z}$, where $\tilde{Z}^T = (\tilde{z}_1, \tilde{z}_2)$, and $\tilde{Z}$ transforms as a doublet under ${\rm SU}(2)_L$ and a singlet under Lorentz rotation.  The complex conjugate $\tilde{Z}^*$ carries $\uonexy$ charge $-1$ and corresponds to a state on the sphere with $-2\pi$ flux in $a_{\mu}$.  These are the same quantum numbers carried by the ${\rm O}(4)$ field $Z$ (and its conjugate $Z^*$), and it is therefore reasonable to identify $\tilde{Z}$ and $Z$.

To work out the transformations of $\tilde{Z}$ under microscopic symmetries, we follow the procedure of Ref.~\onlinecite{hermele08} (which is based on that of Ref.~\onlinecite{alicea05a}), using the transformations for the fermion field $\Phi$ enumerated in Appendix~\ref{app:symms}.  The basic idea is to use the facts that $\tilde{Z}$ is a ${\rm SU}(2)_L$ doublet and carries unit $\uonexy$ charge to write down the most general transformation laws possible, which are
\begin{eqnarray}
T_x : \tilde{Z} &\to& e^{i \phi_T} \exp \Big( \frac{4 \pi i}{3} \sigma^3 \Big) \tilde{Z} \\
R_{\pi/3} : \tilde{Z} &\to& e^{i \phi_R} \sigma^1 \tilde{Z} \\
{\cal R}_y : \tilde{Z} &\to& e^{i \phi_{{\cal R}}} \tilde{Z} \\
{\cal C} : \tilde{Z} &\to& e^{i \phi_c} \sigma^1 \tilde{Z}^* \\
{\cal T} : \tilde{Z} &\to& c_{{\cal T}} \tilde{Z}^* \text{,}
\end{eqnarray}
where $\phi_T$, $\phi_R$, $\phi_{{\cal R}}$ and $\phi_c$ are arbitrary phases, and $c_{{\cal T}} = \pm 1$.  We can partially determine these parameters by demanding that the action of the space group on $\tilde{Z}$ satisfy algebraic relations satisfied by its generators, and by making redefinitions of $\tilde{Z}$.  The relations ${\cal R}_y^2 = R_{\pi/3}^6 = 1$ give the constraints $\phi_{{\cal R}} = 0, \pi$ and $\phi_R = \pi n_R / 3$, where $n_R = 0, 1, \dots, 5$.  Also,  $T_x R^2_{\pi/3} T_x R^{-2}_{\pi/3} = R_{\pi/3} T_x R^{-1}_{\pi/3}$ gives $\phi_T = 0$.  Next, we redefine $\tilde{Z} \to \tilde{Z}' = e^{i \alpha} e^{i \beta \sigma^3} \tilde{Z}$, where $\alpha = - \phi_c / 2$, which has the effect of setting $\phi_c = 0$ in the above transformation laws.  If $n_R < 3$, we choose $\beta = 0$, while for $n_R \geq 3$ we choose $\beta = \pi/2$, which has the effect of sending $n_R \to n_R - 3$ in the transformation laws above.  To summarize, the most general transformation is then characterized by $\phi_T = \phi_c = 0$; $\phi_{{\cal R}} = 0, \pi$; $c_{{\cal T}} = \pm 1$; and $\phi_R = \pi n_R / 3$ where $n_R = 0,1,2$.  If we choose $c_{{\cal T}} = -1$ and $\phi_{{\cal R}} = n_R = 0$, we obtain the same transformation laws as for the ${\rm O}(4)$ field $Z$, so it is indeed consistent with microscopic symmetries to identify $\tilde{Z}$ and $Z$.

{\bf Bilinears $\bI = Z^T (i \sigma^2 \bsigma) Z$ and $\bI^*$.} These objects have $\uonexy$ charge of $\pm 2$.  The QCED3 partners can thus be represented as states on the sphere where $a_{\mu}$ has $4\pi$ flux.  In such a background gauge field, each $\eta_{a \alpha}$ fermion feels $2\pi$ flux, and no flux of the ${\rm SU}(2)$ gauge field is needed to satisfy the Dirac quantization condition.  There are now four zero-energy single particle levels, and the lowest energy monopole state is obtained by filling all negative energy levels and filling the zero-energy levels with two fermions.  We denote the vacuum state for $\pm 4\pi$ flux with no zero-energy levels filled by $| \pm \rangle$, and let $c^\dagger_{a \alpha}$ create a fermion in the zero-energy level corresponding to $\eta_{a \alpha}$.  Focusing on $+ 4\pi$ flux, we consider the class of states $c^\dagger_{a \alpha} c^\dagger_{b \beta} | + \rangle$.  There are a total of 6 such states.  Three of these form a triplet under global ${\rm SU}(2)$ gauge transformations, and thus do not correspond to a local operator.  The other 3 states are a singlet under global ${\rm SU}(2)$ gauge transformations, and a triplet under ${\rm SU}(2)_L$.  These states can be labeled by $| I^i \rangle$ ($i = 1,2,3$) and written 
\begin{equation}
|I^i \rangle = (i \sigma^2)_{\alpha \beta} [\sigma^i  (i \sigma^2) ]_{a b} c^\dagger_{a \alpha} c^\dagger_{b \beta} | + \rangle \text{.}
\end{equation}
  We denote the operator corresponding to these states by the vector $\tilde{\bI}$, while its complex conjugate $\tilde{\bI}^*$ is represented by the corresponding states for $a_{\mu}$ flux of $-4\pi$.

Just as for $\bI$ ($\bI^*$), $\tilde{\bI}$ ($\tilde{\bI}^*$) is a vector under ${\rm SU}(2)_L$, a Lorentz singlet, and carries $\uonexy$ charge of $+2$ ($-2$).  Transformations of $\tilde{\bI}$ under microscopic symmetries can be worked out following the same procedure as above for $\tilde{Z}$, and can be chosen to agree with those of the ${\rm O}(4)$ model operator $\bI$.  Alternatively, if we think of $\tilde{\bI}$ as the composite operator $\tilde{\bI} \sim \tilde{Z}^T (i \sigma^2 \bsigma) \tilde{Z}$, then $\tilde{\bI}$ and $\bI$ transform identically, simply following from the identical transformations of $\tilde{Z}$ and $Z$.

{\bf Conserved current $\bJ^R_{\mu}$.}  We have already identified $2 (\bJ_{\mu}^R)^z$ with the gauge flux $j^G_{\mu}$, but it remains to identify the other components $(\bJ^R_{\mu})^x , (\bJ^R_{\mu})^y$.  We define linear combinations
\begin{equation}
(\bJ^R_{\mu})^{\pm} = (\bJ^R_{\mu})^x \pm i (\bJ^R_{\mu})^y \text{,}
\end{equation}
so that $(\bJ^R_{\mu})^{\pm}$ carries $\uonexy$ charge $\pm 2$ and corresponds to a QCED3 monopole operator.

To construct the corresponding monopole operator using the state-operator correspondence, we let the operators $d^\dagger_{a \alpha \mu p}$ create a fermion in an energy $E_p$ level for $p = \pm 1$.  Here the index $\mu$ transforms in the triplet representation of the Lorentz group.  The states
$\epsilon_{\mu \nu \lambda} d^\dagger_{a \alpha \nu +} \sigma^i_{a b} d^{\vphantom\dagger}_{b \alpha \lambda -} | I^i \rangle$ have the correct quantum numbers to be identified with $(J^R_{\mu})^+$.  However, there is another natural set of candidate states.  Defining the gauge-triplet states
\begin{equation}
|G^i \rangle = (i \sigma^2)_{a b} [\sigma^i  (i \sigma^2) ]_{\alpha \beta} c^\dagger_{a \alpha} c^\dagger_{b \beta} | + \rangle \text{,}
\end{equation}
we see that the states $\epsilon_{\mu \nu \lambda} d^\dagger_{a \alpha \nu +} \sigma^i_{\alpha \beta} d^{\vphantom\dagger}_{a \beta \lambda -} | G^i \rangle$ also have the correct quantum numbers to be identified with $(J^R_{\mu})^+$.  While it is not clear which of these operators should be identified with $(J^R_{\mu})^+$, the more important point is that we have found at least one operator with the correct quantum numbers.

{\bf O(4) ``mass'' term ${\cal O}_r = Z^\dagger Z$.} This object must be identified with a QCED3 operator that is invariant under all symmetries of the field theory.  Moreover, at the O(4) fixed point, it is the most relevant such operator.  Using the QCED3 fixed point as a guide, the most relevant such operators in the large-$N_f$ limit (all with scaling dimension 4) are the Maxwell terms for the two gauge fields, and the following set of three quartic terms:  $(\bar{\Phi} \Phi)^2$, $(\bar{\Phi} \bnu \Phi)^2$ and $(\bar{\Phi} \bmu \Phi)^2$.  These quartic terms are independent, but to form a complete basis for all singlet quartic terms the additional operator $(\bar{\Phi} \gamma_{\mu} \Phi)^2$ must also be included.  However, it can be shown that this operator actually has scaling dimension 6 in the $N_f \to \infty$ limit.\cite{hermele-unpub}  As $N_f$ is reduced from infinity, some of the above dimension-4 operators are expected to become more relevant.  Indeed, to identify the QCED3 and O(4) \emph{fixed points}, one of them must lower its dimension to about 1.67, the dimension of $Z^\dagger Z$ at the ${\rm O}(4)$ critical point, while the others must remain irrelevant (dimensions greater than 3).  To have a clearer indication whether this is the case, it would be useful to calculate $1/N_f$ corrections to the dimensions of these operators.  While similar calculations have appeared in the literature before,\cite{alicea05a, cenkexu08} $N_f = \infty$ shifts in scaling dimension [analogous to the fact that $(\bar{\Phi} \gamma_{\mu} \Phi)^2$ has dimension 6 here] are present but were not taken into account; this will modify the results.  We hope that future work will resolve this issue, which would be useful in a variety of situations where similar field theories arise.

{\bf Anisotropy term ${\cal O}_v = |z_1|^2 |z_2|^2 - (1/6)Z^\dagger Z$.}  This operator is a Lorentz singlet, carries zero $\uonexy$ charge, and belongs to a $\ell_L = 2$ multiplet of ${\rm SU}(2)_L$ (with $m_L = 0$), so, as with ${\cal O}_r$ it is natural to identify it with a QCED3 operator quartic in the fermion fields.  We have classified quartic terms according to their transformations under the Lorentz group and ${\rm SU}(2)_L$, and find that there are two independent terms with the correct quantum numbers:
\begin{eqnarray}
R_0 &=& (\bar{\Phi} \nu^3 \Phi)^2 - \frac{1}{3} (\bar{\Phi} \bnu \Phi)^2 \\
R_1 &=& (\bar{\Phi} \nu^3 \gamma_{\mu} \Phi)^2 - \frac{1}{3}  (\bar{\Phi} \bnu \gamma_{\mu} \Phi)^2 \text{.}
\end{eqnarray}
The O(4) operator ${\cal O}_v$ should be identified with the more relevant of $R_0$ and $R_1$.

{\bf Topological density ${\mathfrak C}$.}  We identify the O(4) operator ${\mathfrak C}$ with the QCED3 bilinear $\tilde{{\mathfrak C}} = -i \bar{\Phi} \Phi$, which is also odd under spatial reflections and time-reversal, but invariant under other symmetries.  

It may seem that this identification spells trouble for a potential duality between O(4) and QCED3 \emph{fixed points}.   ${\mathfrak C}$ is quartic in the ${\rm O}(4)$ field and has three derivatives, thus appearing strongly irrelevant, while $\tilde{{\mathfrak C}}$, as a fermion bilinear, na\"{\i}vely appears likely to be relevant.  However, two points are in order.  First, the O(4) model operator ${\mathfrak C}$ cannot be generalized in a natural way to the ${\rm O}(n)$ case (or to $d=3$), so the above intuition about its scaling dimension is suspect.  Second, the scaling dimension of $\tilde{{\mathfrak C}}$ to leading order in the large-$N_f$ expansion is
\begin{equation}
\operatorname{dim} \tilde{\mathfrak C} = 2 + \frac{256}{3 \pi^2 N_f} + {\cal O}(1/N_f^2) \text{,}
\end{equation}
which, upon inspection of the diagrams involved, can be obtained by simply adding the corresponding anomalous dimensions for QED3\cite{hermele-erratum07} and QCD3.\cite{cenkexu08}  Therefore,  gauge fluctuations make $\tilde{{\mathfrak C}}$ substantially more irrelevant -- the coefficient of $1/N_f$ is quite large.  The striking contrast from the behavior of the $\tilde{\boldsymbol{N}}$ bilinears, which become more relevant with decreasing $N_f$, is interpreted physically in Ref.~\onlinecite{hermele-erratum07}.  It is then plausible that $\tilde{{\mathfrak C}}$ continues to be irrelevant, perhaps strongly so, down to $N_f = 2$.

\section{Evidence for the duality: stable phases}
\label{sec:evidence2}

We shall now show that the three stable phases of the O(4) model Eq.~(\ref{eqn:o4-lagrangian}) -- the paramagnet, and coplanar and collinear ordered states -- are realized simply in QCED3.  While we do not have enough control over QCED3 at $N_f = 2$ to determine the phase diagram, the fact that all the known phases of the O(4) model can be realized is an important piece of evidence in support of the proposed duality.

We begin by considering the paramagnetic phase, which can be represented in the dual description by condensation of vortices.\cite{dasgupta81}  We need to find a vortex condensate that does not break any microscopic symmetries, and to that end we consider
\begin{equation}
V = \Phi^T (i \mu^2) (i \nu^2) (i \tau^2) \Phi \text{,}
\end{equation}
which creates a single vortex.  $V$ is invariant under $T_x$ and ${\cal T}$ and is a scalar under $R_{\pi/3}$.    Under ${\cal R}_y$ and ${\cal C}$, $V \to V^\dagger$.  Working in a fixed gauge, condensation of $V$ means that
$\langle V \rangle = e^{i \theta} | V |  \neq 0$.  If $\theta = 0$, then clearly the condensate preserves the microscopic symmetries.  In fact, even if $\theta \neq 0$, the apparent breaking of ${\cal R}_y$ and ${\cal C}$ is a gauge artifact, as these symmetries can be restored by supplementing them with an appropriate gauge transformation.  Alternatively, one can always choose a gauge such that $\theta = 0$.

We can understand the effect of this vortex condensation on the fermion spectrum, at the mean-field level, by adding the term $(\Delta/2) \int d^2 \br ( V + V^\dagger)$ to the mean-field Hamiltonian.  It is easily seen that the dispersion relation becomes $E(\bk) = \pm \sqrt{\bk^2 + \Delta^2}$, and a gap of $2 \Delta$ is opened.  In this gapped state, the ${\rm SU}(2)$ gauge field will become confining, thus removing single fermion excitations from the spectrum.  Also, the ${\rm U}(1)$ gauge field will be in a Higgs phase, and its photon mode and excitations carrying nonzero $\uonexy$ charge will acquire a gap.  Therefore all excitations are gapped, no symmetries are broken, and we have a description of the paramagnet.

To access the magnetically ordered phases, we consider the Lagrangian ${\cal L} = {\cal L}_{{\rm QCED3}} + \balpha \cdot \tilde{\bN}$.  At the mean-field level (and also at the QCED3 fixed point), the addition of $\balpha \cdot \tilde{\bN}$ will open a gap in the fermion spectrum.  As a result the ${\rm SU}(2)$ gauge field will become confining, and monopole operators of the ${\rm U}(1)$ gauge field will acquire an expectation value -- the latter phenomenon corresponds to the spontaneous breaking of $\uonexy$ symmetry and XY magnetic order.  The photon of the ${\rm U}(1)$ gauge field remains gapless, and corresponds to the spin wave mode of the magnetically ordered state.

To determine the pattern of magnetic order, we note that because $\balpha \cdot \tilde{\bN}$ breaks various space group symmetries, it leads to a gapped ``vortex insulator,'' where the pattern of lattice symmetry breaking depends on $\balpha$.  The vortex insulators that result are the same as those arising in the discussion of Ref.~\onlinecite{alicea05a}, so we give only a brief summary of the results here.  If $\balpha = \alpha_z \bz$, the resulting state is a vortex ``charge density wave,'' where vortices preferentially occupy one of the two honeycomb sublattices.  This corresponds to the $120^\circ$ coplanar magnetically ordered state.\cite{alicea05a}  On the other hand, if $\balpha = \alpha_x \bx + \alpha_y \by$, the resulting state is a vortex ``valence bond solid'' (VBS), where vortices hop back and forth preferentially on a subset of honeycomb lattice bonds.  As detailed in Ref.~\onlinecite{alicea05a}, the vortex VBS states that arise here correspond to the collinear magnetically ordered states.

\section{Discussion}
\label{sec:discussion}

We have given evidence for a rare example of non-abelian duality (without supersymmetry) in $d=2$, between the O(4) model and QCED3 field theories.  Both were derived as low-energy effective descriptions of the same triangular lattice frustrated XY model.  The O(4) model arose from a standard Landau theory treatment.  QCED3 was obtained by combining the abelian boson-vortex duality of the lattice model with a fermionic slave-particle treatment of the vortex degrees of freedom.  By setting up an operator dictionary between the two field theories, and showing that both can realize the three stable phases of the original XY model, we argued that both field theories describe the same low-energy sector of the XY model and are thus dual.

As discussed in Sec.~\ref{sec:statement}, this result is to be understood primarily as a duality between field theories.  It is natural to ask whether it is \emph{also} a duality between the O(4) and QCED3 fixed points, assuming that the QCED3 fixed point exists for $N_f = 2$.  The two possibilities are that either the O(4) and QCED3 fixed points are identical, or they are distinct.  The latter possibility would mean, for example, that the O(4) model realizes both fixed points somewhere in its parameter space.  While we are not aware of any evidence for this, it is hard to rule out, especially if the $N_f = 2$ QCED3 fixed point is highly unstable and requires tuning of several parameters.  However, given the available evidence on scaling dimensions of operators at the two fixed points, it is certainly plausible that they are identical.
Numerical simulations of QCED3 for various values of $N_f$ could potentially shed some light on this issue.

The O(4)-QCED3 duality suggests a number of directions to pursue a further understanding of $d=2$ non-abelian duality.  One can certainly explore whether similar constructions, perhaps starting from other lattice models, lead to other duality relations.  Another natural question is whether there are connections between the present results and $d=2$ dualities of supersymmetric field theories.\cite{intriligator96, kapustin99}  Perhaps upon suitably breaking supersymmetry in a pair of dual theories, the O(4)-QCED3 duality -- or other, related dualities -- can be obtained.

Finally, it is interesting to remark that the O(4)-QCED3 duality can be viewed as a bosonization of QCED3, where the bosonized form is simply the O(4) model.  If more examples of similar dualities are discovered, and better understood, they might eventually be useful as a kind of $d=2$ bosonization.

\acknowledgments{I am grateful to Jason Alicea and T. Senthil for useful comments, and especially to Jason Alicea, Matthew P. A. Fisher  and Olexei Motrunich for a related prior collaboration.}

\appendix

\section{Continuum limit for fermions}
\label{app:fermion-contlimit}

Consider the mean-field Hamiltonian
\begin{equation}
{\cal H}_{{\rm MFT}} = t \sum_{\langle \vr \vr' \rangle} \big( \psi^\dagger_{\vr} \psi^{\vphantom\dagger}_{\vr'} + \text{H.c.} \big) \text{,}
\end{equation}
where the sum is over nearest-neighbor links of the honeycomb lattice.
We will take the continuum limit for the low-energy excitations (which are massless Dirac fermions), and give the relationship between continuum and lattice fields.

We use the 2-site unit cell labeled by $(\vR,i)$, with $i = 1,2$, and define the Fourier transform
\begin{equation}
\psi_{\vR i} = \frac{1}{\sqrt{N_c}} \sum_{\bk} e^{i \bk \cdot \vR} \psi_{\bk i} \text{,}
\end{equation}
where $N_c$ is the number of unit cells in the lattice.  The Hamiltonian then becomes
\begin{equation}
{\cal H}_{{\rm MFT}} = t \sum_{\bk} 
\left( \begin{array}{cc}
\psi^\dagger_{\bk 1} & \psi^\dagger_{\bk 2}
\end{array} \right)
H(\bk) 
\left( \begin{array}{c}
\psi_{\bk 1} \\
\psi_{\bk 2}
\end{array} \right) \text{,}
\end{equation}
where 
\begin{equation}
\left[H(\bk)\right]_{1 1} = \left[H(\bk)\right]_{2 2} = 0 \text{,}
\end{equation}
and
\begin{equation}
\left[H(\bk)\right]_{1 2} = \left[H(\bk)\right]^*_{2 1} = 1 + e^{-i \bk \cdot \ba_2} + e^{i \bk \cdot (\ba_1 - \ba_2)} \text{.}
\end{equation}

The Dirac nodes are at $\bk = \pm \bQ$, where $\bQ = (4\pi/3)\bx$.  Putting $\bk = \pm \bQ + \bq$ and expanding to first order in small $\bq$ we find
\begin{equation}
H(\bQ + \bq) = - \frac{\sqrt{3}}{2} (q_x \tau^1 + q_y \tau^2) \text{,}
\end{equation}
and
\begin{equation}
H(-\bQ + \bq) = \frac{\sqrt{3}}{2} (q_x \tau^1 - q_y \tau^2 ) \text{,}
\end{equation}
where $\tau^i$ are $2 \times 2$ Pauli matrices.
We define the continuum spinors
\begin{equation}
\tilde{\varphi}_1(\bq) \sim \left( \begin{array}{c}
\psi_{\bQ + \bq, 1} \\
\psi_{\bQ + \bq, 2}
\end{array} \right) \text{,}
\end{equation}
and
\begin{equation}
\tilde{\varphi}_2(\bq) \sim \left( \begin{array}{c}
\psi_{-\bQ + \bq, 1} \\
\psi_{-\bQ + \bq, 2}
\end{array} \right) \text{,}
\end{equation}
Note that $\tilde{\varphi}_a$ (with $a = 1,2$) is a 4-component object.  

It is convenient to act on this space with tensor products of two different kinds of Pauli matrices.  The $\tau^i$ Pauli matrices introduced above act in the space labeling the two sites of the unit cell, and the $\mu^i$ Pauli matrices generate $\rSU2$ gauge transformations.  For example, the action of these Pauli matrices on $\tilde{\varphi}_1(\bq)$ is given by
\begin{equation}
\mu^i \tilde{\varphi}_1(\bq) \sim \left( \begin{array}{c}
\mu^i \psi_{\bQ + \bq, 1} \\
\mu^i \psi_{\bQ + \bq, 2}
\end{array} \right) \text{,}
\end{equation}
and
\begin{equation}
\tau^i \tilde{\varphi}_1(\bq) \sim \left( \begin{array}{c}
\left[\tau^i\right]_{1 \alpha}  \psi_{\bQ + \bq, \alpha} \\ 
\left[\tau^i\right]_{2 \alpha}  \psi_{\bQ + \bq, \alpha}
\end{array} \right) \text{,}
\end{equation}
where in each entry there is an implied sum over $\alpha = 1,2$.

The continuum Hamiltonian takes the form
\begin{equation}
{\cal H}_{c} = v \int \frac{d^2\bq}{(2\pi)^2}
\Big[ \tilde{\varphi}^\dagger_1 (-q_x \tau^1 - q_y \tau^2) \tilde{\varphi}_1
+ \tilde{\varphi}^\dagger_2 (q_x \tau^1 - q_y \tau^2) \tilde{\varphi}_2 \Big] \text{,}
\end{equation}
where $v$ is the velocity.  To make this look like the conventional Dirac Hamiltonian we define
\begin{eqnarray}
\varphi_1 &\equiv& e^{i \pi / 6} \tau^3 \tilde{\varphi}_1 \\
\varphi_2 &\equiv& e^{-i \pi / 6} \tau^1 \tilde{\varphi}_2 \text{,}
\end{eqnarray}
and we have
\begin{equation}
{\cal H}_{c} = v \int \frac{d^2\bq}{(2\pi)^2} \varphi^\dagger_a (q_x \tau^1 + q_y \tau^2) \varphi^{\vphantom\dagger}_a \text{.}
\end{equation}

We then define the eight-component object
\begin{equation}
\Phi = \left( \begin{array}{c}
\varphi_1 \\
\varphi_2 
\end{array} \right) \text{.}
\end{equation}
We introduce another set of Pauli matrices, $\nu^i$, acting in this two-component flavor space.  For example,
\begin{equation}
\nu^1 \Phi = \left( \begin{array}{c} \varphi_2 \\ \varphi_1 \end{array} \right) \text{.}
\end{equation}

\section{Symmetries}
\label{app:symms}

Here we enumerate the action of the microscopic symmetries of the triangular lattice XY model Eq.~(\ref{eqn:xymodel}) (listed in Sec.~\ref{sec:directxy}), on both the O(4) model and QCED3 fields.

The action of the symmetries on the O(4) model field $Z^T = (z_1, z_2)$ is easily obtained from $e^{i \phi_{\br}} \sim e^{i \bQ \cdot \br} z_1(\br, \tau) + e^{-i \bQ \cdot \br} z_2(\br, \tau)$, the relationship between lattice and continuum fields discussed in Sec.~\ref{sec:directxy}.  One obtains the following results:
\begin{eqnarray}
T_{x} : Z(\br, \tau) &\to& \exp\Big( \frac{4\pi i \sigma^3}{3} \Big) Z(\br, \tau) \\
R_{\pi/3} : Z(\br, \tau) &\to& \sigma^1 Z(\br', \tau) \\
{\cal R}_y : Z(\br, \tau) &\to& Z(\br', \tau) \\
{\cal C} : Z(\br, \tau) &\to& \sigma^1 Z^*(\br, \tau) \\
{\cal T} : Z(\br, \tau) &\to& - Z^*(\br, \tau) \text{.}
\end{eqnarray}
For $R_{\pi/3}$ and ${\cal R}_y$, $\br'$ is the image of $\br$ under the corresponding symmetry operation.

To obtain the action of the symmetries on the QCED3 fermion field $\Phi$, one first obtains their action on the lattice fermion $\psi_{\bar{r}}$.  The action of each symmetry is chosen so that (1) the vortex operators $v^i_{\bar{r}}$, which are bilinears of $\psi_{\bar{r}}$, have the correct transformation properties, and (2) ${\cal H}_{{\rm MFT}}$ is invariant.  Because $\psi_{\bar{r}}$ is not a gauge-invariant object, the symmetry group acting on it should be thought of as a projective symmetry group.\cite{wen02}  One feature of this situation is that, because ${\cal H}_{{\rm MFT}}$ is invariant under global ${\rm SU}(2)$ gauge transformations, with each symmetry operation we are free to make an arbitrary global ${\rm SU}(2)$ gauge transformation.  The action of the symmetries on $\Phi$ is then easily determined by the relations between continuum and lattice fields given in Appendix~\ref{app:fermion-contlimit}.   Several analyses of this kind have appeared in the literature (see, for example, Ref.~\onlinecite{hermele05}), so we simply quote the results:
\begin{eqnarray}
T_{x} : \Phi(\br) &\to& \exp \Big( \frac{4 \pi i }{3} \nu^3 \Big) \Phi(\br) \\
R_{\pi/3} : \Phi(\br) &\to&   (i \nu^1) e^{i \pi \tau^3 / 6} \Phi(\br') \\
{\cal R}_y : \Phi(\br) &\to& - (i \mu^2) (i \nu^2) \Phi^*(\br') \\
{\cal C} : \Phi &\to&  \tau^1 \nu^1 \Phi^* \\
{\cal T} : \Phi &\to& (i \mu^2) (i \tau^2) (i \nu^2) \Phi \text{.}
\end{eqnarray}

Combined with the requirement that the action be invariant, these transformations determine the transformations of the ${\rm SU}(2)$ gauge field $\alpha^i_{\mu}$.  We also enumerate the transformations of the ${\rm U}(1)$ gauge field $a_{\mu}$.  Under $T_x$, both $\alpha^i_{\mu}$ and $a_{\mu}$ are invariant.  Under $R_{\pi/3}$, both $a_0$ and $\alpha^i_{0}$ transform as scalars, while the spatial components rotate as vectors.  Under the remaining symmetries,
\begin{eqnarray}
{\cal R}_y : (a_0, a_1, a_2) &\to& (-a_0, -a_1, a_2) \\
{\cal R}_y : (\alpha^i_0, \alpha^i_1, \alpha^i_2) &\to& (\alpha^i_0, \alpha^i_1, - \alpha^i_2) \\
{\cal C}: a_\mu &\to& - a_{\mu} \\
{\cal C} : (\alpha^1_{\mu}, \alpha^2_{\mu}, \alpha^3_{\mu}) &\to& (- \alpha^1_{\mu}, \alpha^2_{\mu}, -\alpha^3_{\mu})  \\
{\cal T} : a_{\mu} &\to& - a_{\mu} \\
{\cal T} : \alpha^i_{\mu} &\to& \alpha^i_{\mu} \text{.}
\end{eqnarray}
Note that the transformations of $\alpha^i_{\mu}$ depend on the arbitrary choice of global ${\rm SU}(2)$ gauge transformation in the corresponding transformation of $\Phi$.

\section{Non-vanishing of $2\pi$-flux monopole state}
\label{app:nonzero}

Here we show that $| \psi \rangle$, the gauge-invariant $2\pi$-flux monopole state constructed in Eq.~(\ref{eqn:ginvt-monopole}) from $| \psi_0 \rangle$, the gauge-fixed state, is nonzero.  The Hilbert space of the theory on the sphere is a tensor product of fermion and gauge field Hilbert spaces, and we can write
\begin{equation}
| \psi_0 \rangle = | f_0 \rangle \otimes | \tilde{a}_{\mu}(\Omega) , \tilde{\alpha}^i_{0, \mu}(\Omega) \rangle \text{,}
\end{equation}
where $| f_0 \rangle$ gives the state of the fermions, and the latter factor the state of the gauge fields in the vector potential basis.  $\Omega = (\theta, \phi)$ specifies the angular position on the sphere.  The vector potential states of the gauge field Hilbert space form an orthonormal basis.  In spherical coordinates, the vector potentials above are given by $\tilde{a}_{\theta} = \tilde{\alpha}^i_{\theta} = 0$ and
\begin{eqnarray}
\tilde{a}_{\phi} &=& \frac{1 - \cos \theta}{2 \sin \theta} \\
\tilde{\alpha}^i_{\phi} \sigma^i &=& \frac{1 - \cos \theta}{2 \sin \theta} \sigma^3 \text{.}
\end{eqnarray}
This corresponds to a monopole with $2\pi$ flux piercing the sphere in both $a_{\mu}$ and $\alpha^3_{\mu}$.

To show that $| \psi \rangle \neq 0$, it is enough to show that $\langle \psi_0 | \psi \rangle \neq 0$.  Consider a general gauge transformation $G = G_1 G_2$ acting on $| \psi_0 \rangle$, where $G_1$ is a general ${\rm U}(1)$ gauge transformation and $G_2$ a general ${\rm SU}(2)$ gauge transformation.  Below, we show that either 
\begin{equation}
G | \psi_0 \rangle = | \psi_0 \rangle \text{,}
\end{equation}
or
\begin{equation}
G | \psi_0 \rangle = | f \rangle \otimes | \bar{a}_{\mu}(\Omega), \bar{\alpha}^i_{0, \mu}(\Omega) \rangle \text{,}
\end{equation}
 where $\bar{a}_{\mu} \neq \tilde{a}_{\mu}$ or $\bar{\alpha}^i_{\mu} \neq \tilde{\alpha}^i_{\mu}$.  In the latter case, $\langle \psi_0 | G | \psi_0 \rangle = 0$.  This implies that
\begin{equation}
\label{eqn:psi-form}
| \psi \rangle =  \int [d G_1] \int [d G_2] G_1 G_2 | \psi_0 \rangle = C | \psi_0 \rangle + | \psi' \rangle \text{,}
\end{equation}
where $C > 0$ and $\langle \psi_0 | \psi' \rangle = 0$.  Therefore $\langle \psi_0 | \psi \rangle = C \neq 0$, and $| \psi \rangle$ is indeed nonzero.

It is clear that $| \psi \rangle$ would be zero if either the $a_{\mu}$ or $\alpha^3_{\mu}$ charge were nonzero.  The reason is that the integral of Eq.~(\ref{eqn:psi-form}) is a projection onto gauge-singlet states, if either the $a_{\mu}$ or $\alpha^3_{\mu}$ charge is nonzero, this projection must vanish.  This shows that it is necessary for these charges to vanish in order to have nonzero $| \psi \rangle$.

Let $s(\Omega)$ be the generator of ${\rm U}(1)$ gauge transformations, and $t_i(\Omega)$ ($i = 1,2,3$) the generators of ${\rm SU}(2)$ gauge transformations.  Then a general ${\rm U}(1)$ gauge transformation can be written $G_1 = \exp(i \int d\Omega \lambda(\Omega) s(\Omega))$ and a general ${\rm SU}(2)$ gauge transformation is $G_2 = \exp( i \int d\Omega \lambda_i(\Omega) t_i(\Omega) )$.  Because $| \psi_0 \rangle$ has zero charge under $a_{\mu}$ and $\alpha^3_{\mu}$, it is invariant under global ${\rm U}(1)$ gauge transformations ($\lambda(\Omega)$ constant), and a global ${\rm SU}(2)$ gauge transformation where $\lambda_1 = \lambda_2 = 0$ and $\lambda_3(\Omega)$ is constant.

It is easy to see that any global ${\rm SU}(2)$ gauge transformation other than that above changes $\tilde{\alpha}^i_{\mu}$, so it rotates $| \psi_0 \rangle$ into a new state orthogonal to $| \psi_0 \rangle$.  The same is true for any non-global ${\rm U}(1)$ gauge transformation ($\lambda(\Omega)$ not constant), because all such transformations change $\tilde{a}_{\mu}$.  We now show that all non-global ${\rm SU}(2)$ gauge transformations change $\tilde{\alpha}^i_{\mu}$, which implies the result we needed to show (\emph{i.e.} that $G$ either leaves $|\psi_0\rangle$ invariant or rotates it into an orthogonal state).

Under an ${\rm SU}(2)$ gauge transformation, the object $A_{\mu}(\Omega) = \tilde{\alpha}_{\mu}^i(\Omega) \sigma^i$ transforms as
\begin{equation}
A_{\mu} \to U A_{\mu} U^\dagger + i U \partial_{\mu} U^\dagger \text{,}
\end{equation}
where $U = U(\Omega) =  \exp(i \lambda^i(\Omega) \sigma^i)$.  In order for $\tilde{\alpha}^i_{\mu}$ to be invariant, the above relation must become an equality for both $\mu = \theta$ and $\mu = \phi$ components.  Since $A_{\theta} = 0$, we therefore must have $\partial_{\theta} U = 0$, and $U$ is independent of $\theta$.  The $\mu = \phi$ equation gives
\begin{equation}
i U \partial_{\phi} U^\dagger = \frac{1 - \cos\theta}{2 \sin\theta} \Big[ \sigma^3 - U \sigma^3 U^\dagger \Big] \text{.}
\end{equation}
Since the term in brackets on the right-hand is independent of $\theta$, and the left-hand side must be independent of $\theta$, the only consistent possibility is that both $\partial_{\phi} U = 0$, which implies that $\lambda_i$ is constant and the gauge transformation is global, and also  $U \sigma^3 U^\dagger = \sigma^3$, which implies $\lambda_1 = \lambda_2 = 0$.  Therefore, there are indeed no ${\rm SU}(2)$ gauge transformations leaving $\tilde{\alpha}^i_{\mu}$ invariant other than that identified above.

\bibliography{duality-paper}

\end{document}